\documentclass[aps,showpacs,twocolumn]{revtex4}
\usepackage{graphicx}

\newcommand{\be}{\begin{eqnarray}}
\newcommand{\ee}{\end{eqnarray}}
\newcommand{\ba}{\begin{array}}
\newcommand{\ea}{\end{array}}
\newcommand{\bmat}{\left(\begin{array}}
\newcommand{\emat}{\end{array}\right)}
\newcommand{\bw}{\begin{widetext}}
\newcommand{\ew}{\end{widetext}}
\newcommand{\no}{\nonumber}
\newcommand{\e}{\epsilon}
\newcommand{\tr}{\mbox{tr}\,}
\newcommand{\str}{\mbox{str}\,}

\begin{document}
\title{Chaotic scattering through coupled cavities}
\author{Kazutaka Takahashi}
\affiliation{Department of Physics, Tokyo Institute of Technology,  
 Tokyo 152--8551, Japan}
\author{Tomosuke Aono}
\affiliation{Department of Physics, Ben--Gurion University of the Negev,
 Beer--Sheva 84105, Israel}
\date{\today}

\begin{abstract}
 We study the chaotic scattering 
 through an Aharonov-Bohm ring containing two cavities.
 One of the cavities has well-separated resonant levels while
 the other is chaotic, and is treated by random matrix theory.
 The conductance through the ring is calculated analytically 
 using the supersymmetry method and 
 the quantum fluctuation effects are numerically investigated in detail.
 We find that the conductance is determined by the competition between 
 the mean and fluctuation parts.
 The dephasing effect acts on the fluctuation part only. 
 The Breit-Wigner resonant peak is changed to 
 an antiresonance by increasing the ratio of the level broadening 
 to the mean level spacing of the random cavity, 
 and the asymmetric Fano form turns into a symmetric one.
 For the orthogonal and symplectic ensembles, the period of 
 the Aharonov-Bohm oscillations is half of that for regular systems.
 The conductance distribution function becomes independent of 
 the ensembles at the resonant point, which can be understood by 
 the mode-locking mechanism.
 We also discuss the relation of our results 
 to the random walk problem.
\end{abstract}
\maketitle

\section{Introduction}

 Starting from the study of atomic nuclei,
 chaotic scattering has been a topic of intensive research 
 in a large variety of systems such as 
 atoms, molecules, quantum devices, and microwave cavities
 \cite{CS}.
 A fundamental question to be asked 
 is how much information is reflected 
 in the scattering through random media such as 
 disordered and classically chaotic systems.
 One of the most remarkable and promising ideas 
 is to introduce the statistical concept into the analysis.
 The ensemble average over different realizations of the sample
 is considered to calculate several statistical quantities.
 A large number of systems exhibit universal behavior 
 determined by the symmetry of the systems.
 For this situation, random matrix theory (RMT) \cite{WD,Mehta,GMW} 
 has been used to understand the result and 
 has played an important role as a standard analytical tool.

 Recently, the experimental stage of the chaotic scattering 
 has been shifted from natural to artificial systems.
 Typical examples are mesoscopic systems
 \cite{MRWHG,Beenakker,Alhassid,Hacken,ABG} 
 such as quantum dots (QDs) and disordered wires.
 Recent development of nanotechnology makes it possible to 
 fabricate mesoscopic quantum hybrid systems that could not be realized
 before and a lot of interesting interference phenomena have been
 observed under controllable external parameters.
 Due to the interference of wave functions, 
 a system made from parts such as the QD, lead, and quantum point contact
 cannot be treated separately.
 Such systems show new interesting phenomena
 which are absent in single isolated systems.
 Typical experimentally fabricated systems are
 the QD on the Aharonov-Bohm (AB) ring \cite{YHMS}, 
 the side-coupled QD \cite{KASKI}, and so on.

 The model treated in this paper is 
 two QDs put on the two arms of the AB ring.
 In this so-called ``mesoscopic double slit system,'' 
 a lot of interesting phenomena such as the AB oscillations 
 and the Fano effects can be observed by the interference of 
 wave functions transmitting through the two arms 
 \cite{GIA,KK,KAKI,NTA}.
 In the context of chaotic scattering, 
 it is interesting to apply the known analysis based on RMT 
 \cite{VWZ,pker,PEI,ISS,randomS,BB94,Brouwer,FS}
 to the AB ring system.
 We study how the interference effects appear and 
 the conductance behaves as the function of the controllable 
 parameter such as the magnetic flux through the ring.

 Our formulation is rather general and 
 the application of our result is not limited to the QD systems.
 It is known that microwaves in an irregular shaped cavity 
 behave chaotically and the statistical properties can be 
 described by RMT \cite{Stockmann,HZOAA}.
 Based on the formal analogy between the Helmholtz and 
 Schr\"odinger equations, the classical waves are simulated 
 as quantum mechanical wave functions.
 Compared with the mesoscopic systems in nanoscale, 
 the cavity system is easier to fabricate 
 and is ideal for an experimental study.
 We can also observe the Fano effect in this system \cite{RLBKS}.

 How can we define the statistical model for the hybrid system?
 For the system of two QDs attached to each arm of the AB ring,
 Gefen {\it et al.} \cite{GIA} considered the case when 
 each dot has a single regular level.
 As a simple but nontrivial extension, 
 we treat the case when one of the dots 
 has regular levels and the other has random levels.
 RMT is applied to the random dot.

 This model can be viewed as a mixed system of 
 chaotic and integrable levels.
 In single dot systems, such structure 
 is employed as an idea to explain anomalous phenomena 
 such as critical statistics \cite{KLAA} and 
 fractal conductance \cite{micolich}.
 It is known in the open QD system that 
 the several specific levels couple with the lead strongly while 
 the other levels couple weakly via strong coupled levels \cite{SI}.
 Thus it is too simple to treat the dot as a single random matrix
 and we need to consider the internal structure more seriously.
 Although our model is not directly related to such phenomena, 
 it is instructive and useful to consider 
 the present ring system as the situation 
 where the strong and weak couplings coexist.
 In this system, the regular transmission in the one arm is 
 affected by the random ones in the other arm, and vice versa.

 From a point view of RMT,
 special attention is paid to the universality of 
 the statistical quantities.
 A natural question to be asked in the present model 
 is how the universal level correlations described by RMT 
 are modified by the regular contribution.
 Naive expectation is that the effect of the regular levels 
 can be safely removed by the proper scaling (unfolding) \cite{BHZ}.
 It is known that the effective theory is written in terms 
 not of the microscopic parameters but of
 the transmission coefficients \cite{VWZ}. 
 However, in the present system, 
 the effect is amplified by multiple scatterings 
 through the ring and gives highly nontrivial results.

 Now that our model has been described, 
 we must refer to the work by Clerk {\it et al.}\cite{CWB}.
 They considered many resonant levels in a single dot and 
 RMT was employed for their distribution.
 The regular component to the S matrix 
 expressing the direct nonresonant path through the dot 
 was used to find the Fano resonances.
 For each resonance, the Fano parameter $q$ was calculated and 
 the statistical distribution of $q$ was defined over the resonances.
 On the other hand, in our case, 
 only the single resonant level is present regularly
 and it is affected by random levels.
 Thus our attention is fixed on the single regular resonance.
 To discuss the statistical properties of the transport 
 we must prepare different realizations of the random dot.
 The ensemble average is defined in terms of such realizations.

 The outline of this paper is the following.
 The AB ring model is defined in Sec.\ref{model}.
 We define the random Hamiltonian model in Sec.\ref{randomH}.
 A model based on the random S matrix is also defined in
 Sec.\ref{randomS} 
 and the relation to the random Hamiltonian model is discussed.
 In Sec.\ref{avS}, we calculate the average of the S matrix 
 based on the random Hamiltonian model.
 As a result the mean part of the conductance is calculated.
 It is not enough to calculate the conductance including 
 the quantum fluctuation effect and 
 we develop the supersymmetry method \cite{Efetov} in Sec.\ref{susy}
 to calculate the full conductance.
 The results of the conductance are shown in Sec.\ref{conductance}.
 We also study the AB oscillations in Sec.\ref{ABO}
 and the Fano effect in Sec.\ref{fanoeff}.
 The fluctuation effects can be best seen 
 in the conductance distribution functions, 
 which are studied in Sec.\ref{cdf}.
 Since realistic situations are not ideal
 and phase breaking effect is present \cite{HZOAA,HPMBDH}, 
 it is important to consider the dephasing effect theoretically.
 We consider it in Sec.\ref{dephasing}
 using a simple imaginary-potential model.
 Section \ref{conc} is devoted to discussions and conclusions.
 Part of the results were published in a preliminary report \cite{TA}.

\section{Model}
\label{model}

\subsection{Random Hamiltonian approach}
\label{randomH}

\begin{figure}[tb]
\begin{center}
\includegraphics[width=0.9\columnwidth]{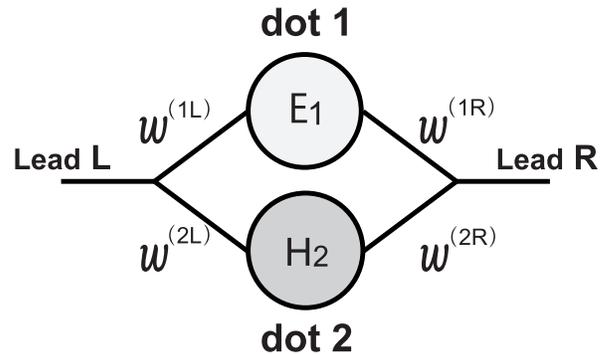}
\caption{Schematic drawing of our model.
 Dot 1 with a resonant level $E_1$ and 
 dot 2 with a random Hamiltonian $H_2$
 are connected by leads.
 $w$ denotes a dot-lead coupling matrix.}
\label{abring}
\end{center}
\end{figure}

 We consider the AB ring system depicted in Fig.\ref{abring}.
 The upper dot (dot 1) has a single resonant level, and 
 the lower dot (dot 2) has random levels and is
 treated by RMT.
 It is known from scattering theory that the S matrix 
 of the system is written as \cite{Beenakker,Alhassid,VWZ,FS}
\be
 S &=& 1-2\pi i w^\dag\frac{1}{E-H+i\pi ww^\dag}w \no \\
 &=& \frac{1-i\pi w^\dag\frac{1}{E^+-H}w}{1+i\pi w^\dag\frac{1}{E^+-H}w},
 \label{S}
\ee
 where $H$ denotes the Hamiltonian matrix for dots 
 and $w$ the dot-lead coupling matrix.
 $H$ can be written as 
\be
 H = \bmat{cc} E_1 & 0 \\ 0 & H_2 \emat,
\ee
 where $E_1$ is the fixed energy level for the dot 1  
 and $H_2$ is the random Hamiltonian for the dot 2.
 The size of $H_2$, $N$, is taken to be infinity 
 to find the universal result.
 We note that the total size of $H$ is $1+N$.
 It is a straightforward task to extend the size of 
 the upper dot Hamiltonian to arbitrary values
 and here we consider the minimal size 1.
 As another simplification, 
 we consider the $2\times 2$ (unitary) matrix $S$, which means
 that the left and right leads have a single channel, respectively.
 It is believed that the quantum interference effect becomes 
 maximal in this case \cite{BB97}.
 Then the dot-lead coupling matrix $w$
 is the $(1+N)\times 2$ matrix and can be written as 
\be
 w &=& \bmat{cc} w^{(1)} \\ w^{(2)} \emat
 = \bmat{cc} w^{(1L)} & w^{(1R)} \\ w^{(2L)} & w^{(2R)} \emat \no\\
 &=& \bmat{cc} w^{(1L)} & w^{(1R)} \\ w^{(2L)}_1 & w^{(2R)}_1 \\
 w^{(2L)}_2 & w^{(2R)}_2 \\  
 \vdots & \vdots \\
 w^{(2L)}_N & w^{(2R)}_N
 \emat,
\ee
 where $(1L)$ refers to the coupling between 
 the dot $1$ and lead $L$, and so on.

 The conductance measures the transmission from the left to right lead 
 and is defined by \cite{Beenakker, Alhassid, VWZ}
\be
 g = \langle |S_{12}|^2\rangle, \label{cond}
\ee
 where $\langle\ \rangle$ denotes the ensemble averaging of 
 the random Hamiltonian $H_2$.
 We employ the Gaussian ensemble \cite{Mehta} and 
 the probability density is given by 
\be
 P(H) = C\exp\left(-\frac{\pi^2}{2N\Delta^2}\tr H^{2}\right), 
\ee
 where $\Delta$ is the mean level spacing, and 
 $C$ is the normalization constant.
 In the following calculations, 
 we mainly consider unitary symmetry, which means that 
 $H_2$ is Hermitian and no additional condition is imposed.

 The result of the conductance depends on the choice of 
 the dot-lead coupling $w$.
 Although this matrix $w$ has $4N$ degrees of freedom, 
 there is no need to specify them completely.
 After the averaging, the effective degrees of freedom becomes finite.
 Generally, it is 6 and 
 we restrict our discussion to the special case of 4 (see below).

\subsection{Random S matrix approach}
\label{randomS}

 Equation (\ref{S}) is a useful formula to relate
 the Hamiltonian to the S matrix 
 and can be used for the present coupled system.
 It is convenient to express the S matrix in terms of the K matrix 
 defined by 
\be
 S = \frac{1-iK}{1+iK}.
\ee
 $K$ is expressed as the sum of contributions from dot 1 and 2:
\be
 & & K = K_1+K_2, \no \\
 & & K = \pi w^\dag\frac{1}{E^+-H}w, \no\\
 & & K_1=\frac{\pi w^{(1)\dag}w^{(1)}}{E^+-E_1}, \; 
 K_2=\pi w^{(2)\dag}\frac{1}{E^+-H_2}w^{(2)}. \label{K}
\ee
 This simple relation implies the sum rule of the S matrix 
\be
 \frac{1-S}{1+S} = \frac{1-S_1}{1+S_1}+\frac{1-S_2}{1+S_2},
 \label{sr}
\ee
 where $S_1$ ($S_2$) is the S matrix for dot 1 (2).
 It is instructive and useful 
 in the following numerical calculations
 to derive the explicit representation using the matrix elements.
 Defining each S matrix elements as 
\be
 S = \bmat{cc} r & t' \\ t & r' \emat, \;
 S_i = \bmat{cc} r_i & t_i' \\ t_i & r_i' \emat \; (i=1,2),
 \label{SS1S2}
\ee
 we obtain, for example, 
\be
 t &=& 4\Bigl[t_1(1+s_2+r_2+r_2')
 +t_2(1+s_1+r_1+r_1')\Bigr] \no\\
 & & \times \Bigl[
 9+3(r_1+r'_1+r_2+r'_2)+s_1+s_2 \no\\
 & & -3(r_1r_2+r'_1r'_2)+(r_1r'_2+r_2r'_1)
 -4(t_1t'_2+t_2t'_1) \no\\
 & & 
 -(r_1+r'_1)s_2-(r_2+r'_2)s_1+s_1s_2
 \Bigr]^{-1}, \label{t} 
\ee
 where $s_i=\det S_i=r_ir'_i-t_it'_i$ ($i=1,2$).
 Thus the total transmission $t$ is not equal to 
 $t_1+t_2$, rather including nonlinear effects
 due to multiple scattering through the ring.
 Such multiple scattering effects are put together with 
 interference due to random scattering
 and give highly nontrivial results for the conductance
 $g=\langle|t|^2\rangle$.

 Another way of representing the total S matrix is 
 to separate the S matrix of the system into 
 the upper and lower dot parts and the left and right fork parts
 \cite{GIA, KK}.
 Choosing the fork matrices in a proper way, 
 we can find the same expression of $t$ as in Eq.(\ref{t}).

 The conductance can be calculated by taking the ensemble average 
 over $S_2$ determined by the random Hamiltonian $H_2$.
 Instead of doing that, 
 we may disregard the detailed structure of $S_2$ 
 and impose randomness directly on $S_2$,
 simulated by the circular ensembles \cite{Mehta}.
 It is well known that the random S matrix approach is equivalent to 
 the random Hamiltonian approach if we use 
 the Poisson kernel \cite{pker} 
\be
 P_\beta(S)d\mu_\beta(S) \propto 
 \frac{1}
 {\left|\det\left(1-S\langle S\rangle^\dag
 \right)\right|^{2\beta+2-\beta}}
 d\mu_\beta(S), \label{PK}
\ee
 where $d\mu_\beta(S)$ denotes the invariant measure for the S matrix 
 and is used as the measure for the circular ensemble.
 $\beta$ is the index for the universality class.
 $\beta=1,2,$ and 4 for the unitary, orthogonal, and symplectic case, 
 respectively. 
 $\langle S\rangle$ is the averaged value of S which is
 treated as an input parameter and 
 is determined by the random Hamiltonian model.
 The total S matrix is constructed by the sum rule (\ref{sr})
 and the conductance is expressed by $|t|^2$ where 
 $t$ is given by Eq.(\ref{t}).
 By taking the circular ensemble average with 
 the weight $P_\beta(S_2)$, 
 we obtain the conductance $g$ which is the same as 
 that obtained by the random Hamiltonian approach.
 The equivalence of both approaches was shown in Ref.\cite{Brouwer}.
 The random S matrix approach has a great advantage 
 for numerical calculations
 because there is no need to take the thermodynamic limit $N\to\infty$
 and one may consider $2\times 2$ random matrices $S_2$.

 Alternatively, we can parametrize the S matrix 
 in terms of the K matrix (\ref{K}).
 Then the expression of the conductance becomes 
 much simpler than Eq.(\ref{t}) as we show in Sec.\ref{cdf}.
 The disadvantage of this parametrization 
 is that the K matrix is Hermitian and the matrix elements are not compact,
 which is inconvenient for the numerical calculation.
 Thus we employ the S matrix parametrization (\ref{t}) with 
 compact variables for most of the numerical calculations.

\section{Averaged S matrix}
\label{avS}

 As we have shown in Eq.(\ref{K}), 
 the K matrix is written as the sum of the regular (dot 1) and
 random (dot 2) parts.
 Thus, to get the averaged K matrix, 
 we may consider the ensemble averaging of the random part.
 We know from RMT that the averaged Green function 
 for the Gaussian unitary ensemble is given by \cite{Mehta}
\be
 \left<\frac{1}{E^+-H_2}\right> = \frac{\pi}{N\Delta}e^{-iz},
\ee
 where 
\be
 \cos z = \frac{\pi E}{2N\Delta}.
\ee
 $N$ is taken to be infinity while $E/\Delta$ is kept finite.
 Then we have $e^{-iz}\to -i$ and 
 the averaged K matrix is given by
\be
 \langle K\rangle = \frac{1}{E^+-E_1}\gamma_1
 -\frac{i\pi}{N\Delta}\gamma_2,
\ee
 where $\gamma_i=\pi w^{(i)\dag}w^{(i)}$ ($i=1,2$).
 It is important to note that the result depends
 on the dot-lead couplings $w^{(1,2)}$ through $\gamma_{1,2}$.

 For the regular dot, the most general form of $\gamma_1$ is
\be
 \gamma_1 &=& \bmat{cc} \pi w^{(1L)*}w^{(1L)} & \pi w^{(1L)*}w^{(1R)} \\
 \pi w^{(1R)*}w^{(1L)} & \pi w^{(1R)*}w^{(1R)} \emat \no\\
 &=& \frac{1}{2}\bmat{cc} \Gamma_{1L} & \sqrt{\Gamma_{1L}\Gamma_{1R}}e^{-i\phi} \\ 
 \sqrt{\Gamma_{1R}\Gamma_{1L}}e^{i\phi} & \Gamma_{1R} \emat,
\ee
 where $\Gamma_{1L}$ ($\Gamma_{1R}$) turns out to be the level width for 
 the left (right) coupling of the dot to the lead and 
 $\phi$ is a phase.
 We assume the symmetric coupling $\Gamma_{1L}=\Gamma_{1R}$ 
 for simplicity and use 
\be
 \gamma_1 = \Gamma_1\Phi,
 \label{gamma1}
\ee
 where 
\be
 \Phi= \frac{1}{2}\bmat{cc} 1 & e^{-i\phi} \\ e^{i\phi} & 1 \emat.
\ee
 This matrix satisfies $\Phi^2=\Phi$ and 
 is diagonalized as $\Phi\to{\rm diag} (0,1)$.

 On the other hand, for the random dot, 
 the form of $\gamma_2$ is slightly complicated.
 It is written as
\be
 \gamma_2 = \bmat{cc} \pi w^{(2L)\dag}w^{(2L)} & \pi w^{(2L)\dag}w^{(2R)} \\
 \pi w^{(2R)\dag}w^{(2L)} & \pi w^{(2R)\dag}w^{(2R)} \emat.
\ee
 Since $w^{(2L)}$ and $w^{(2R)}$ are $N\times 1$ matrices, 
 we see that the relation 
 $|w^{(2L)\dag}w^{(2L)}||w^{(2R)\dag}w^{(2R)}|\ge
 |w^{(2L)\dag}w^{(2R)}||w^{(2R)\dag}w^{(2L)}|$
 holds.
 The equal sign holds when $w^{(2L)}=w^{(2R)}$ or 
 $N=1$, 
 the latter is the case for $w^{(1)}$.
 Thus we need the additional parameter 
 for the parametrization of $\gamma_2$.
 Assuming the symmetry of the left and right coupling again, 
 we obtain the form 
 with the level width $\Gamma_2$ as 
\be
 \gamma_2 = 
 \frac{N\Gamma_2}{2}\bmat{cc} 1 & ae^{i\phi} \\ ae^{-i\phi} & 1 \emat.
 \label{gamma2}
\ee
 The parameter $a$ reflects the above mentioned inequality and 
 $0\le a\le 1$.
 We note that the same phase $\phi$ appears in $\gamma_1$ and $\gamma_2$, 
 but the sign is opposite to each other.
 This phase affects the transmission part of the S matrix and 
 can be identified with the AB flux through the ring \cite{GIA,KK}.

 Using this parametrization, we can write
\be
 \langle K\rangle = \frac{1}{\e}\Phi
 -\frac{iX}{2}\bmat{cc} 1 & ae^{i\phi} \\ ae^{-i\phi} & 1 \emat,
 \label{avK}
\ee
 where 
\be
 \e = \frac{E-E_1}{\Gamma_1}, \;
 X=\frac{\pi\Gamma_2}{\Delta}.
\ee
 The energy $\e$ represents the distance 
 from the resonance point and
 $X$ is the ratio of the level width to the mean level spacing of the dot 2.
 Thus this model is described by four parameters 
 $\e$, $X$, $a$, and $\phi$.

 For the random dot, 
 the elements of the dot-lead coupling $w^{(2)}$ 
 distribute randomly 
 and the summation $\sum_{i=1}^Nw^{(2L)*}_iw^{(2R)}_i$ can be small 
 when the random phases of $w^{(2L)}$ and $w^{(2R)}$ almost cancel out.
 This means $a$ is vanishingly small.
 On the other hand, the summation can be finite 
 when the left and right dot-lead couplings are correlated mutually.
 This results in direct nonresonant reaction \cite{VWZ}.
 We first discuss the case of $a=0$ for simplicity.
 The averaged K matrix takes the form
\be
 \langle K\rangle = \frac{1}{\e}\Phi-\frac{iX}{2}.
\ee
 The finite-$a$ effect is discussed afterwards.

 Now we go back to the S matrix.
 The averaged S matrix is simply obtained 
 by using the averaged K matrix, 
\be
 \langle S\rangle 
 &=& \frac{1-\langle K\rangle}{1+\langle K\rangle} \no\\
 &=& \frac{1-\frac{X}{2}}{1+\frac{X}{2}}
 -\frac{2i}{\left(1+\frac{X}{2}\right)
 \left[\left(1+\frac{X}{2}\right)\e+i\right]}\Phi. \label{Scl}
\ee
 This is justified by the saddle-point analysis of 
 the nonlinear sigma model described below.
 We define $g_0 =|\langle S_{12}\rangle|^2$, 
 which is the conductance if we can disregard the quantum fluctuations.
 It is given by
\be
 g_0 
 = \frac{1}{\left(1+\frac{X}{2}\right)^2}
 \frac{\left(\frac{\Gamma_1}{1+\frac{X}{2}}\right)^2}
 {(E-E_1)^2+\left(\frac{\Gamma_1}{1+\frac{X}{2}}\right)^2}. \label{gcl}
\ee
 The result shows that 
 the level width $\Gamma_1$ for the dot 1 
 and the conductance are renormalized by the factor $1/(1+X/2)$.

 For later use, we define the transmission coefficients as
\be
 T &=& 1-\langle S\rangle\langle S\rangle^\dag \no\\
 &=& \frac{2X}{\left(1+\frac{X}{2}\right)^2}
 -\frac{2X}{\left(1+\frac{X}{2}\right)^2}
 \frac{\left(\frac{\Gamma_2}{1+\frac{X}{2}}\right)^2}
 {(E-E_1)^2
 +\left(\frac{\Gamma_2}{1+\frac{X}{2}}\right)^2}\Phi. \no\\
 \label{T}
\ee
 This matrix can be diagonalized to find the eigenvalues 
\be
 T_1 = \frac{2X}{\left(1+\frac{X}{2}\right)^2}, \;
 T_2 = \frac{2X}{\left(1+\frac{X}{2}\right)^2+\frac{1}{\e^2}}.
 \label{T12}
\ee
 Note that $ 0 \leq T_2 \leq T_1$,
 $T_2=T_1$ at $|E-E_1|\to\infty$, and 
 $T_2=0$ at $E=E_1$.
 At $X=2$, $T_1$ takes the maximum value, $T_1=1$, and 
 the transmission through the random dot becomes ideal. 

 In conclusion of this section, 
 we found the averaged S matrix (\ref{Scl}) 
 and the conductance (\ref{gcl}).
 Of course, this is not the final result of the averaged conductance.
 We just calculated the mean part $g_0=|\langle S_{12}\rangle|^2$
 which is different from the original definition (\ref{cond}).
 We must examine the fluctuation part 
 $\delta g= g-g_0=\langle |S_{12}|^2\rangle-|\langle S_{12}\rangle|^2$.

\section{Conductance}

\subsection{Supersymmetry method}
\label{susy}

 We derive the nonlinear sigma model for the coupled system
 to calculate the fluctuation part of the conductance.
 According to the supersymmetry method \cite{Efetov, VWZ}, 
 the generating function for the product of Green functions 
 $G(E)=1/(E-H+i\pi ww^\dag)$ and $G^\dag(E)$ is defined by 
\be
 Z = \int {\cal D}(\bar{\psi},\psi)
 \exp\Bigl[i\bar{\psi}
 \Bigl(E+i\Lambda\pi ww^\dag-H\Bigr)\psi\Bigr],
\ee
 where $\psi$ has $4(1+N)$ components coming from supersymmetry
 (bosons and fermions), 
 retarded-advanced structure, and Hamiltonian space.
 $\Lambda={\rm diag} (1,-1)$ in retarded-advanced space.
 Following the standard procedure, 
 we introduce the Hubbard-Stratonovitch field $Q$
 to write the averaged generating function as 
\be
 \left<Z\right> 
 &=& \int {\cal D}Q 
 \exp\biggl\{
 -\str_{4(1+N)}\ln\biggl[E+i\Lambda\pi ww^\dag \no\\
 & & -\bmat{cc}
 E_1 & 0 \\
 0 & \frac{N\Delta}{\pi} Q \emat\biggr]
 -\frac{1}{2}\str_4 Q^2\biggr\},
\ee
 where $Q$ is a 4$\times$4 supermatrix.
 ``$\str$'' denotes supertrace and 
 the subscript indicates the size of superspace. 
 When $w=0$, the saddle-point equation is written down as
\be
 Q=\frac{N\Delta}{\pi}\frac{1}{E^+-\frac{N\Delta}{\pi}Q}.
\ee
 This is easily solved with the proper boundary condition as 
\be
 Q =  e^{-iz\Lambda}
 = \frac{\pi E}{2N\Delta}
 -i\Lambda\sqrt{1-\left(\frac{\pi E}{2N\Delta}\right)^2} 
 \to -i\Lambda,
\ee
 where we took the limit $N\to\infty$ keeping $E/\Delta$ finite.
 As a general solution including the saddle-point manifold, 
 we can write 
\be
 Q = -i\sigma, \; 
  \sigma = V\Lambda \bar{V},
\ee
 where $V$ is the $4\times 4$ supermatrix and satisfies $V\bar{V}=1$.
 The symmetry of $V$ is determined 
 in the standard way \cite{Efetov}.

 Now the generating function reads 
\be
 \left<Z\right> 
 &=& \int {\cal D}\sigma e^{-F}, \no\\
 F &=& \str_{4(1+N)}\ln\left[
 1+\bmat{cc} \frac{1}{E-E_1} & 0 \\
 0 & \frac{\pi}{N\Delta}Q \emat i\Lambda \pi ww^\dag \right] \no\\
 &=& \str_8\ln\left[
 1+i\Lambda\left(
 \frac{1}{E-E_1}\gamma_1
 -\frac{i\pi}{N\Delta}\gamma_2\sigma\right) 
 \right],
\ee
 where we assumed $\gamma_1=\Gamma_1\Phi$ and $\gamma_2=N\Gamma_2/2$.
 Since the matrix sizes of $\sigma$ and 
 $\gamma$ are 4 and 2, respectively,
 the total size of the superspace in the last expression is 8.
 We finally obtain 
\be
 F &=& \str_8\ln\left[
 1+i\left(
 \frac{1}{\e}\Phi
 -\frac{iX}{2}\Lambda\right)\sigma
 \right]
 \no\\
 &=& \frac{1}{2}\str_8\ln\left[
 1+\frac{\frac{T}{2}}{1-\frac{T}{2}}
 \frac{\Lambda\sigma+\sigma\Lambda}{2}
 \right]. \label{nls}
\ee
 The nonlinear sigma model (\ref{nls}) 
 can be written in terms of the transmission matrix $T$ (\ref{T})
 and the microscopic fundamental parameter $X$ 
 does not appear in the expression explicitly.
 This is a manifestation of the universality \cite{Mehta,GMW,VWZ}.

 Equation (\ref{nls}) is for systems with unitary symmetry.
 In the same way we can derive the nonlinear sigma models 
 for the orthogonal and symplectic symmetry classes.
 Then $\sigma$ becomes an $8\times 8$ supermatrix and 
 the additional symmetry due to time-reversal invariance is 
 imposed \cite{Efetov}.

\subsection{Conductance}
\label{conductance}

 In the nonlinear sigma model approach,
 the averaged conductance is calculated by 
 performing the integration of the $\sigma$ matrix.
 The mean part of the conductance $g_0$ in Eq.(\ref{gcl})
 is easily obtained by neglecting the fluctuation of 
 the $\sigma$ matrix as $\sigma=\Lambda$.
 To find the fluctuation part of the conductance 
 $\delta g=\langle |S_{12}|^2\rangle-|\langle S_{12}\rangle|^2$, 
 we must take into account the contribution from 
 the saddle-point manifold parametrized by the $V$ matrix.
 This calculation is highly complicated and 
 we refer to the Appendix \ref{calcond} for details.
 We finally obtain 
\be
 \delta g &=& \frac{T_1+T_2}{4}
 -\left(\frac{1-\frac{X^2}{4}}{2X}\right)
 \left(\frac{1-\frac{X^2}{4}-\frac{1}{\e^2}}{2X}\right) \no\\
 & & \times\left(\frac{T_1T_2}{T_1-T_2}\right)^2
 \left[\frac{T_1+T_2}{2}
 -\left(\frac{T_1T_2}{T_1-T_2}\right)\ln \frac{T_1}{T_2}
 \right], \no\\
 \label{g1}
\ee
 where $T_{1,2}$ are the eigenvalues of the transmission matrix $T$
 given by Eq.(\ref{T12}).

\begin{figure}[tb]
\begin{center}
\includegraphics[width=0.9\columnwidth]{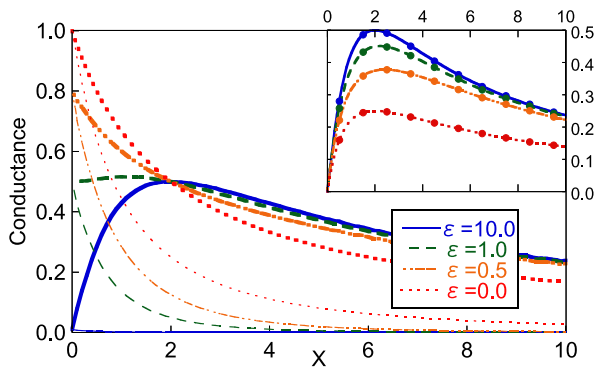}
\caption{Conductance vs $X=\pi\Gamma_2/\Delta$. 
 The thick (thin) lines are 
 analytical results of the total conductance $g$ 
 (mean part $g_0$).
 Inset : Comparison of the analytical (denoted by lines) 
 and numerical (dots) 
 results for the fluctuation part $\delta g=g-g_0$.
}
\label{gx}
\end{center}
\end{figure}

 We first examine the two limiting cases, $|\e|\to\infty$ and $\e=0$.
 The limit $|\e|\to\infty$ means that dot 1 is detached
 from the system and the S matrix is given by $S_1=1$.
 In this case, $T_1=T_2=2X/(1+X/2)^2$ and 
 we recover the known result \cite{Efetov2} 
\be
 \delta g = \frac{T_1}{3}+\frac{T_1^2}{6}. \label{g1single}
\ee
 In the other limit $\e=0$ ($E=E_1$) 
 the energy coincides with the level in dot 1 and 
 the perfect transmission through dot 1 is achieved.
 Then $T_2=0$ and we obtain 
\be
 \delta g = \frac{T_1}{4}. \label{g1res}
\ee
 We see that Eq.(\ref{g1single}) is larger than Eq.(\ref{g1res}),
 which means that the fluctuation effects are reduced 
 as we approach the resonant point.
 For intermediate values of $\e$,
 Eq.(\ref{g1}) cannot be written in terms of $T_{1,2}$ only 
 in contrast to Eqs.(\ref{g1single}) and (\ref{g1res}).
 This is because the source term to calculate the conductance 
 depends on $\e$ and $X$ explicitly,
 although the nonlinear sigma model itself can be written in terms of $T$,
 as shown in the Appendix \ref{calcond}.
 
 These results are checked by numerical calculations.
 We use the formula (\ref{t}) for the transmission matrix.
 $S_1$ is given by $S_1=(1-iK_1)/(1+iK_1)$ with $K_1=\Phi/\e$, 
 and the random S matrix $S_2$ is treated statistically 
 by using the Poisson kernel (\ref{PK}).
 We take the ensemble average over more than 
 $10^6$ samples of the S matrix.

 In Fig.~\ref{gx}, $X$ dependence of the conductance 
 is shown for several values of $\e$.
 $g_0$ shows a peak at $X=0$
 while $\delta g$ takes a maximum at $X=2$ as shown by 
 the thin lines and the inset in Fig.~\ref{gx}, respectively.
 As $\e\to\infty$ 
 $g_0$ ($\delta g$) is monotonically decreasing (increasing) 
 and the result rapidly approaches Eq.(\ref{g1single}).
 The numerical result agrees with Eq.(\ref{g1}) 
 in a highly accurate way,
 which shows the equivalence of the random Hamiltonian and
 random S matrix approach.

\begin{figure}[tb]
\begin{center}
\includegraphics[width=0.9\columnwidth]{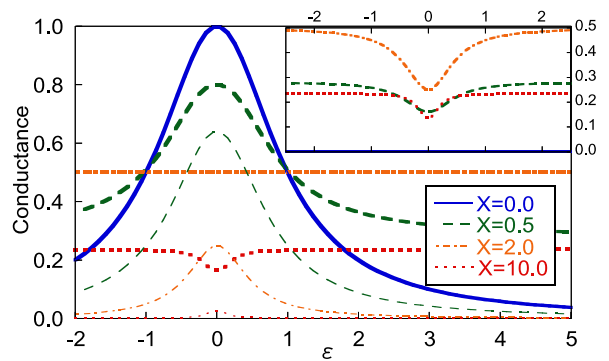}
\caption{Conductance vs $\e=(E-E_1)/\Gamma_1$. 
 The thick (thin) lines are the total conductance $g$ 
 (the mean part $g_0$).
 Inset : Fluctuation part $\delta g=g-g_0$.
}
\label{ge}
\end{center}
\end{figure}

\begin{figure}[tb]
\begin{center}
\includegraphics[width=0.9\columnwidth]{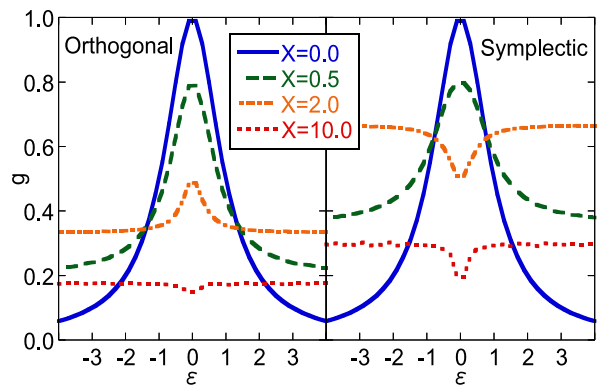}
\caption{Numerical results of the conductance $g(\e)$ 
 for orthogonal and symplectic ensembles.
 The result for the symplectic case is normalized to unity. 
}
\label{oseg}
\end{center}
\end{figure}

 $\e$ dependence of the conductance is shown in Fig.~\ref{ge}.
 A resonance peak appears at $\e=0$, 
 reflecting transport through the regular dot 1.
 This peak structure, however, changes qualitatively as a function of $X$.
 For small $X$ the peak is convex and 
 the peak height decreases on increasing $X$.
 When $X=2$, $g$ is independent of $\e$.
 Increasing $X$ further, we find that the peak turns into 
 an antiresonance and $g$ decreases monotonically.
 The result for $X=2$ corresponds to that of 
 the circular unitary ensemble because $\langle S_2\rangle=0$, and 
 the Poisson kernel $P_\beta(S_2)$ becomes unity.
 As we see in the inset of Fig.\ref{ge}, 
 $\delta g$ at the resonant point is relatively small and 
 the quantum fluctuation effect smooths the resonance.

 For comparison we calculate $g$ as a function of $\e$ for
 the orthogonal and symplectic ensembles numerically.
 For the orthogonal case, 
 the Hamiltonian has time-reversal invariance and 
 the matrix elements are real.
 For symplectic, 
 the Hamiltonian becomes a quaternion real matrix \cite{Mehta}.
 The results are shown in Fig.\ref{oseg}.
 When $0<X<2$, the resonance is enhanced (reduced) for
 the orthogonal (symplectic) ensembles.
 At $X=2$, the orthogonal ensemble gives a resonance while
 the symplectic ensemble gives an antiresonance.
 When $X=10$, we see that antiresonance is reduced (enhanced) 
 for the orthogonal (symplectic) ensemble in contrast to the case of $X<2$.
 
 Away from the resonance, the quantum fluctuation effect 
 becomes larger 
 as the number of degrees of freedom of random variables increases.
 We note that the number of degrees becomes maximum when $\beta=4$
 and minimum when $\beta=1$.
 We can also see that the conductance at the resonant point 
 is independent of the choice of the ensemble.
 This result is discussed in detail in Sec.\ref{cdf}.
 
\subsection{Aharonov--Bohm oscillations}
\label{ABO}

 For regular ring systems, 
 it is well known that 
 the AB oscillations are observed by 
 applying the magnetic flux through the ring.
 Since the flux is a tunable parameter 
 it is an important method to control the system.
 Our interest is how the effect of the AB flux  
 can be observed in the present random system.
 Can the AB oscillations survive after the random averaging?

\begin{figure}[tb]
\begin{center}
\includegraphics[width=0.9\columnwidth]{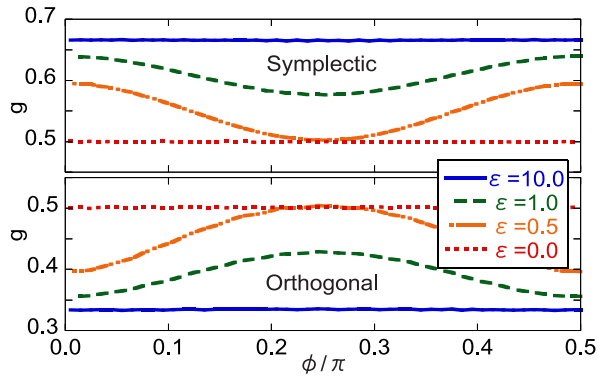}
\caption{Conductance vs $\phi$ for $X=2$.
 The lower figure is for the orthogonal ensemble and 
 upper for symplectic.
 No oscillations are observed for the unitary case. 
}
\label{ab}
\end{center}
\end{figure}

 In systems with unitary symmetry, 
 since the scattering in the random dot randomizes 
 the phase of the amplitude, 
 the result becomes independent of the AB phase $\phi$.
 This is not the case for the orthogonal and symplectic systems 
 and the oscillations can be observed.
 However, the period of the oscillation 
 is different from that for regular systems.
 This can be understood from the expression of 
 the transmission $t$ in Eq.(\ref{t}).
 The phase is included in that expression as 
\be
 t = \frac{At_1e^{i\phi}+Bt_2e^{-i\phi}}
 {C-D(t_1t'_2e^{2i\phi}+t_2t'_1e^{-2i\phi})},
\ee
 where $A$, $B$, $C$, and $D$ are phase independent contributions.
 If we neglect the multiple scattering effect 
 the total transmission is approximated by 
 $t\sim t_1e^{i\phi}+t_2e^{-i\phi}$. 
 Then the conductance is given by 
\be
 g &\sim& |t_1e^{i\phi}+t_2e^{-i\phi}|^2 \no\\
 &=& |t_1|^2+|t_2|^2+t_1t_2^*e^{2i\phi}+t_1^*t_2e^{-2i\phi}.
\ee
 We see that the third and fourth terms of the right hand side
 give oscillations with the period $\pi$.
 However, these terms vanish after the random averaging.
 The contributions going around the ring twice 
 give oscillations with the period $\pi/2$ and 
 survive after the averaging.
 Such contributions come from expanding the denominator.
 Thus, in the orthogonal and symplectic systems, 
 $g$ depends on the AB phase due to
 the multiple scattering inside the ring.
 The period of the AB oscillations becomes half of that
 for the regular systems.
 This effect can be interpreted as
 a kind of the Altshuler-Aronov-Spivak effect \cite{AAS} 
 for cylinder systems.
 In a ring system, it was discussed in Ref.\cite{AAS2} that 
 the period of the oscillation becomes half a flux quantum
 by the self-averaging effect.
 
 We show the numerical results in Fig.\ref{ab} for 
 the orthogonal and symplectic ensembles.
 The period of the oscillations is $\pi/2$ as we discussed above
 and the difference between these two results is that 
 the conductance becomes minimum (maximum) for orthogonal 
 (symplectic) at $\phi=0$.
 This can be understood by the standard mechanism of 
 weak localization \cite{weakl}.

\subsection{Fano effect}
\label{fanoeff}

 The Fano effect is induced by the correlation of 
 the resonant and direct path \cite{CWB,Fano}. 
 The direct path can be described by 
 the parameter $a$ in Eq.(\ref{gamma2}).
 If we keep this parameter in Eq.(\ref{avK}), 
 the averaged S matrix is given by 
\bw
\be
 \langle S\rangle &=& \frac{1}
 {\left(1+\frac{1+a}{2}X\right)\left(1+\frac{1-a}{2}X\right)
 +\left(1+\frac{1-a\cos 2\phi}{2}X\right)\frac{i}{\e}} \no\\
 & & \times\bmat{cc}
 1-\frac{1-a^2}{4}X^2-\frac{1-a\cos 2\phi}{2}X\frac{i}{\e} & 
 -\frac{i}{\e} e^{-i\phi}-a X e^{i\phi} \\
 -\frac{i}{\e} e^{i\phi}-a X e^{-i\phi} &
 1-\frac{1-a^2}{4}X^2-\frac{1-a \cos 2\phi}{2}X\frac{i}{\e}  \emat, 
\ee
\ew
 The mean part of the conductance is derived from this expression as 
\be
 g_0 &=& \frac{a^2 X^2}{\left(1+\frac{1+a}{2}X\right)^2
 \left(1+\frac{1-a}{2}X\right)^2} \no\\
 & & \times \frac{\left|\e-\e_1+q\Gamma_1\right|^2}
 {(E-E_1)^2
 +\frac{\left(1+\frac{1-a\cos 2\phi}{2}X\right)^2}
 {\left(1+\frac{1+a}{2}X\right)^2
 \left(1+\frac{1-a}{2}X\right)^2}\Gamma_1^2}, \label{g0fano}
\ee
 where $q$ is the Fano parameter 
\be
 q = \frac{i e^{2i\phi}}{a X}. \label{fanoq}
\ee
 Thus the Fano effect appears when $a\ne 0$.
 The additional condition $\phi\ne 0$ is required 
 to obtain a finite real part of $q$.
 Then the asymmetric conductance form is obtained.
 At the limit $|E-E_1|\to\infty$, $g_0$ has a finite contribution  
 in contrast with Eq.(\ref{gcl}).
 This means that there is a direct regular coupling between 
 the left and right leads through the random dot.

\begin{figure}[tb]
\begin{center}
\includegraphics[width=0.9\columnwidth]{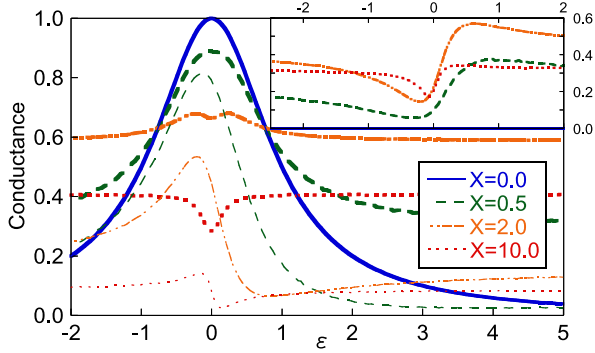}
\caption{Conductance vs $\e=(E-E_1)/\Gamma_1$
 at $a=0.7$ and $\phi=-\pi/8$. 
 The thick (thin) lines are the total conductance $g$ 
 (the mean part $g_0$).
 Inset : Fluctuation part $\delta g=g-g_0$.
 The total conductance $g$ is obtained numerically and 
 the mean part $g_0$ is plotted by using Eq.(\ref{g0fano})
}
\label{fanog}
\end{center}
\end{figure}

 This Fano effect also appears on $\delta g$.
 The numerical result in Fig.\ref{fanog} 
 shows that the Fano parameter for $\delta g$ is 
 the same as Eq.(\ref{fanoq}).
 Since the antiresonance appears in $\delta g$ 
 as shown in the inset of Fig.\ref{ge}, 
 the asymmetry is opposite to that of $g_0$.
 As a result, the total conductance becomes symmetric.
 This result means that the Fano effect appears 
 not on $g$ but on the mean part $g_0$ and 
 the fluctuation part $\delta g$, respectively.
 We note that the asymmetric form is obtained when ${\rm Re}\,q\ne 0$.
 The effect of the imaginary part of $q$ keeps the conductance symmetric.
 We can conclude that the real part of the Fano parameter does not 
 affect the total conductance.
 We confirmed that the symmetric conductance is obtained 
 for the orthogonal and symplectic classes as well.

\section{Conductance distribution functions --- mode-locking effect}
\label{cdf}

 In the previous section we focused on the averaged conductance.
 It is well known that disordered systems show 
 strong fluctuation effects, 
 which mean that the square of the conductance
 and even the higher moments 
 become relevant to characterize the system.
 To discuss the effects of the fluctuations,
 here we calculate the conductance distribution 
 function $P(g)=\langle\delta(g-|S_{12}|^2)\rangle$.
 We show the analytical results when $\e=0$ 
 which show universality among the ensembles. 
 We also report the numerical results.

\begin{figure}[tb]
\begin{center}
\includegraphics[width=0.75\columnwidth]{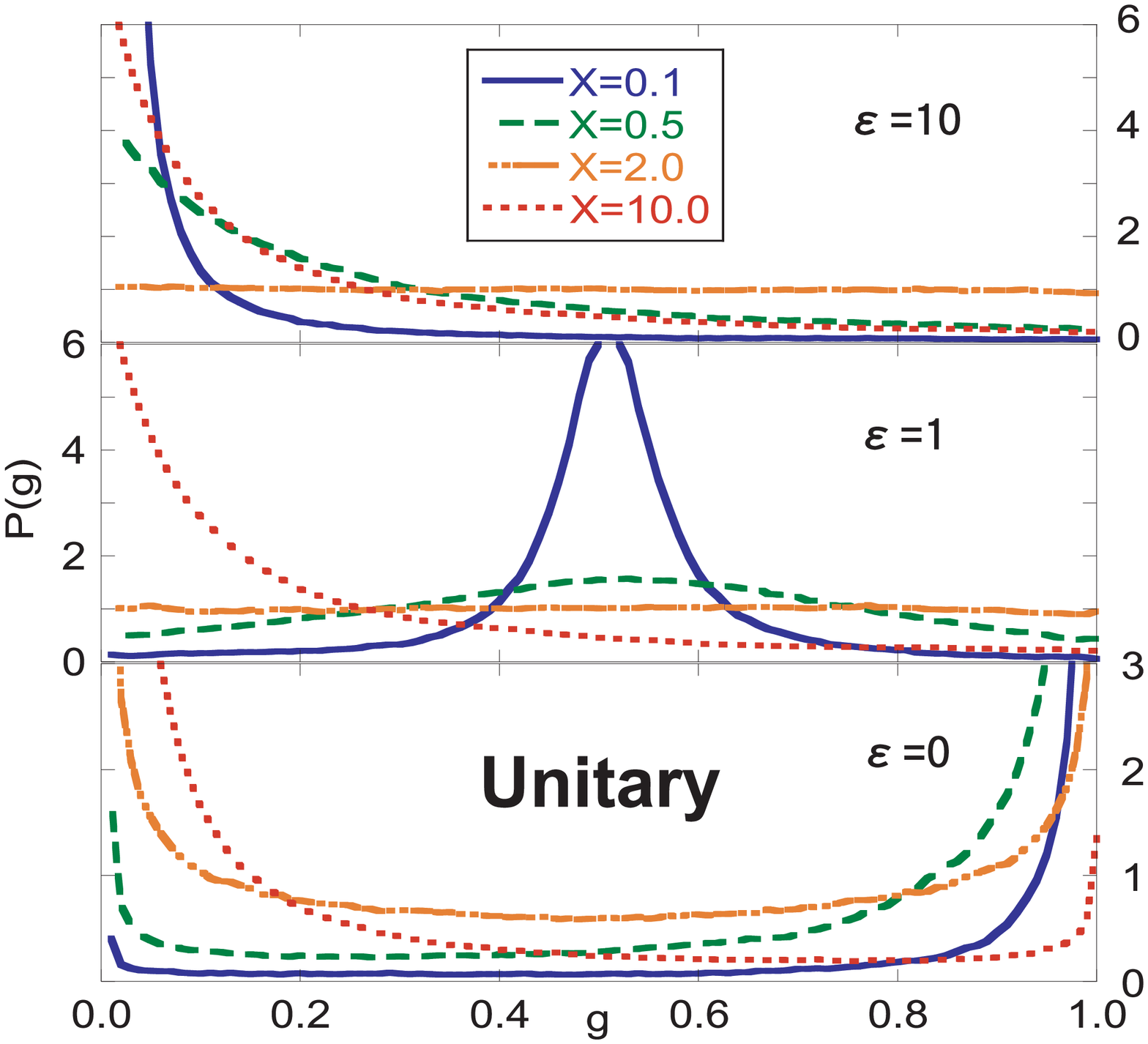}
\includegraphics[width=0.75\columnwidth]{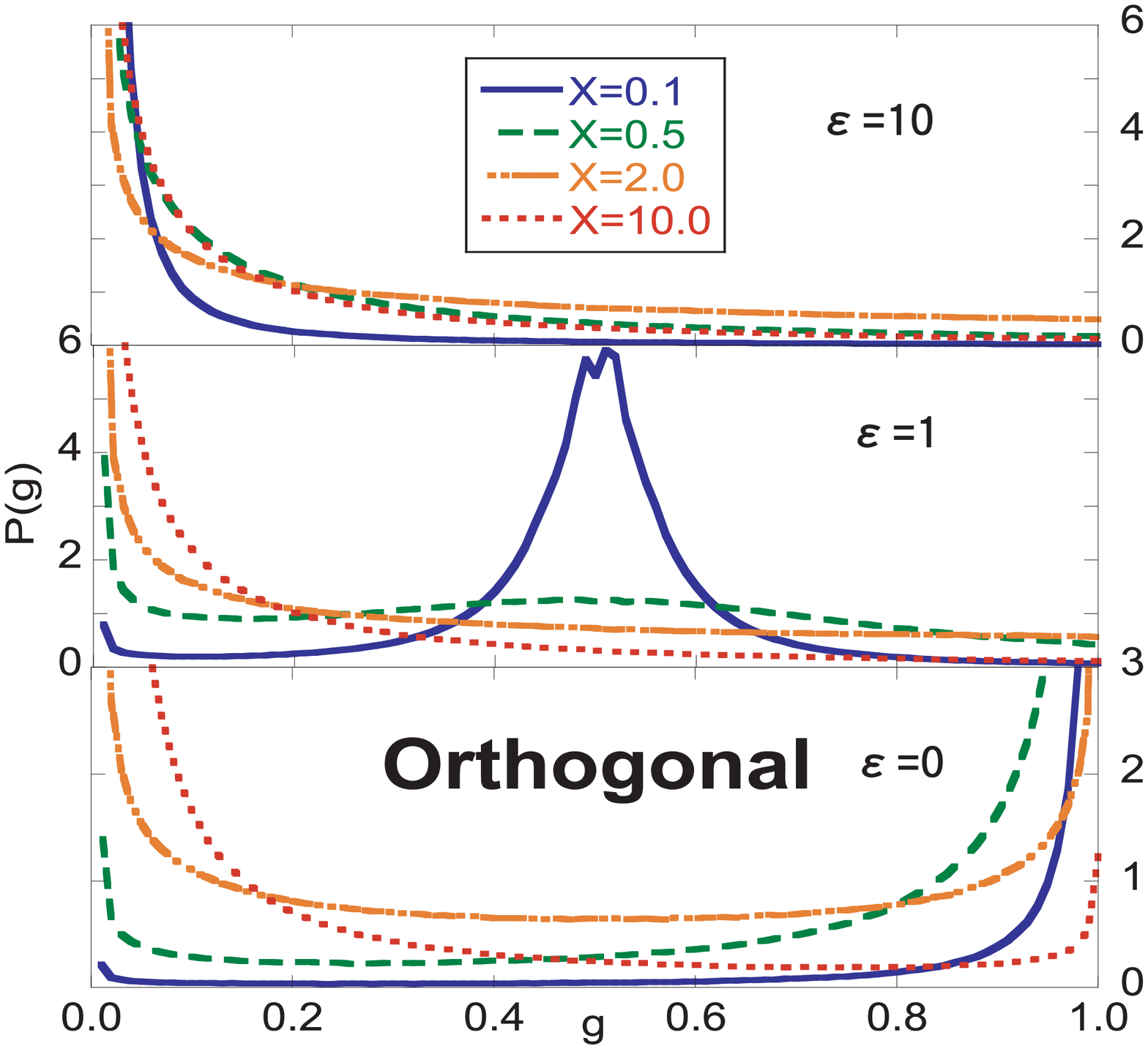}
\includegraphics[width=0.75\columnwidth]{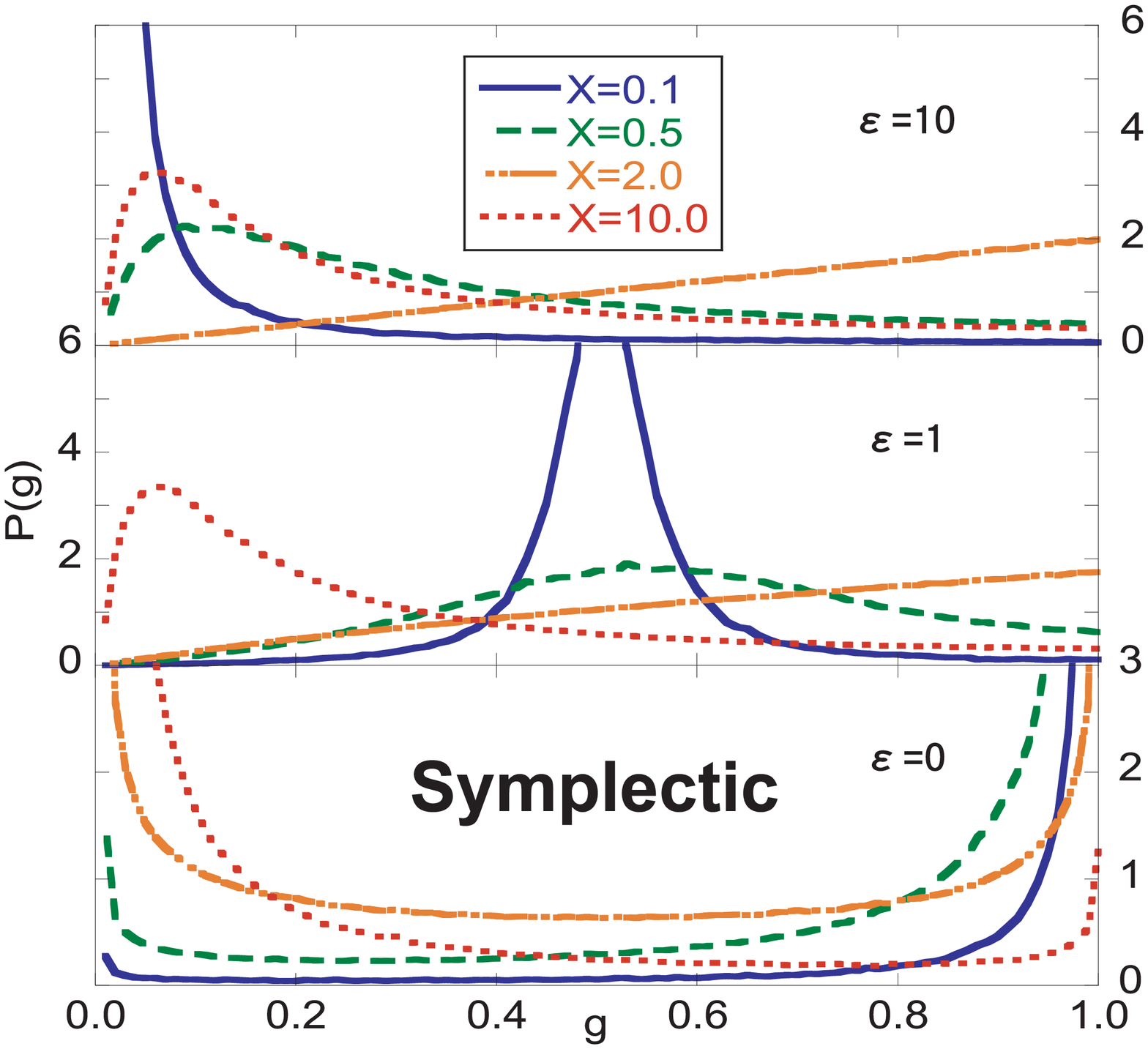}
\caption{Ensemble dependence of the conductance distribution 
functions at $\phi=0$.
The curves at $\e=0$ are well fitted by the analytical result (\ref{pg0})}
\label{pg}
\end{center}
\end{figure}

 The expression of the conductance distribution becomes simpler 
 if we use the K matrix representation 
 as we mentioned in Sec.\ref{randomS}.
 $K$ is a Hermite matrix and $K = K_1 + K_2$ with
\be
 K_1 = \frac{1}{2\e}\bmat{cc}1 & e^{-i\phi} \\ e^{i\phi} & 1 \emat, \;
 K_2 = \bmat{cc} a_1 & be^{i\phi} \\ b^\dag e^{-i\phi} & a_2 \emat.
 \label{Kpara}
\ee
 $a_{1,2}$ are real, and 
 $b$ depends on the universality class 
 and is expressed as
\be
 b = \left\{\ba{cc}
 b_0 & \mbox{for}\ \beta=1, \\
 b_0+ib_1 & \mbox{for}\ \beta=2, \\
 b_0+b_1e_1+b_2e_2+b_3e_3 & \mbox{for}\ \beta=4,
 \ea\right.
\ee
 where $b_{0,1,2,3}$ are real and 
 $e_{1,2,3}$ quaternion matrices defined by 
 $e_j=i\sigma_j$ with the Pauli matrix $\sigma_j$ ($j=1,2,3$).
 The conductance is expressed in this parametrization as 
\bw
\be
 g = \left\{\ba{cc}
 \frac{1+4b_0\e\cos 2\phi+4|b|^2\e^2}
 {\left[b_0\cos 2\phi-\frac{a_1+a_2}{2}+(1-a_1a_2+|b|^2)\e\right]^2
 +\left[1+(a_1+a_2)\e\right]^2} 
 & \mbox{for}\ \beta=1, \\
 \frac{1+4(b_0\e\cos 2\phi-b_1\sin 2\phi)+4|b|^2\e^2}
 {\left[b_0\cos 2\phi-b_1\sin 2\phi
 -\frac{a_1+a_2}{2}+(1-a_1a_2+|b|^2)\e\right]^2
 +\left[1+(a_1+a_2)\e\right]^2} 
 & \mbox{for}\ \beta=2, \\
 \frac{1}{2}\tr\frac{1+4b_0\e\cos 2\phi+4|b|^2\e^2}
 {\left[(b_0\cos 2\phi-r\sigma_3\sin 2\phi
 )-\frac{a_1+a_2}{2}+(1-a_1a_2+|b|^2)\e\right]^2
 +\left[1+(a_1+a_2)\e\right]^2} 
 & \mbox{for}\ \beta=4,
 \ea\right.
\ee
 where $|b|^2=\sum_{i=0}^{\beta-1} b_i^2$ 
 and $r^2= b_1^2+b_2^2+b_3^2$.
 We note that the conductance for $\beta=4$ is normalized to unity.
 This expression is averaged by
 the generalized circular ensemble
 (Poisson kernel)
\be
 P_\beta (S_2)d\mu_\beta(S_2) \propto 
 \left\{\frac{1}{\left[\frac{X}{2}+\frac{2}{X}(-a_1a_2+|b|^2)\right]^2
 +(a_1+a_2)^2}\right\}^{(\beta+2)/2}
 da_1da_2\prod_{i=0}^{\beta-1}db_i, 
\ee
\ew
 where we used $\langle S_2\rangle=(1-X/2)/(1+X/2)$.
 The numerical results using the Metropolis algorithm \cite{Heermann}
 are shown in Fig.\ref{pg} at $\phi=0$.

 The results at large $\e$ are interpreted as the single random dot case.
 This case was discussed in Ref.\cite{PEI} 
 using the random Hamiltonian approach
 and the analytical result for the unitary system was obtained.
 In the random S matrix approach, 
 the case of the perfect transmission $X=2$ was 
 obtained in Ref.\cite{randomS} as
\be
 P(g)=\frac{\beta}{2}g^{-1+\beta/2}, 
\ee
 and other cases were discussed in Ref.\cite{BB94}.
 The case of $\e=10$ is enough to find a large-$\e$ result
 and we find a good agreement with their results.

 In the case of $\e=1$,  
 we can clearly see how the random dot significantly 
 affects the distribution function.
 If we increase $X$, 
 a single peak at small $X$ turns into a broad one 
 and a different peak around $g=0$ is formed at large $X$.

 It is interesting to see the results at $\e=0$ 
 which are independent of the choice of the ensemble.
 In this case,
 the conductance distribution can be calculated analytically.
 The conductance is written as
\be
 g = \frac{1}{1+\left(b_0-\frac{a_1+a_2}{2}\right)^2}. \label{gb0}
\ee
 and the distribution function is obtained from the expression 
\be
 & & P(g) 
 = C\int da_1 da_2\prod_{i=0}^{\beta-1}db_i \no\\
 & & \times
 \delta\left(g-\frac{1}{1+\left(b_0-\frac{a_1+a_2}{2}\right)^2}\right)\no\\
 & & \times
 \left\{\frac{1}{\left[\frac{X}{2}+\frac{2}{X}(-a_1a_2+|b|^2)\right]^2
 +(a_1+a_2)^2}\right\}^{(\beta+2)/2}, \no\\
\ee
 where $C$ is the normalization constant.
 We perform the integrals and obtain 
\be
 P(g) = \frac{1}{\pi\sqrt{g(1-g)}}
 \frac{1}{\frac{2}{X}(1-g)+\frac{X}{2}g}. \label{pg0}
\ee
 This result agrees with the numerical ones 
 in Fig.\ref{pg}.
 The reason why this result becomes independent of $\beta$
 can be considered as follows.
 In Eq.(\ref{Kpara}),
 all the matrix elements of $K_1$ are divergent at $a=0$.
 When $\phi=0$, this diverging term belongs to the member of 
 the orthogonal ensemble and 
 affects the variables $a_{1,2}$ and $b_0$ in the second term $K_2$
 which are common to all the ensembles.
 Then the effective modes are locked on those for the orthogonal class 
 and the conductance (\ref{gb0}) becomes independent of 
 the rest of the parameters $b_{1,2,3}$.

 Equation (\ref{pg0}) with $X=2$ appears in the problem of 
 the classical random walk \cite{feller}
 and is known as the arcsine law.
 Consider the one-dimensional classical random walk starting at the origin.
 The walker can move to either one of its two nearest neighbor sites
 with the equal probability $p=0.5$.
 After the $N$-step walk, 
 we count the number of the events which 
 the walker was in the positive axis $M$.
 Then the distribution function of $g=M/N$ approaches 
 Eq.(\ref{pg0}) with $X=2$ as $N\to\infty$.
 It can be considered that 
 the walker at the positive (negative) direction 
 corresponds to the transmission (reflection) 
 to the left (right) lead in our model.
 Due to the presence of the resonant path through the dot 1,
 a particle transmitted through the dot 2 
 can go to either left or right lead with equal probability.
 The particle reflected by the dot 2 can go either way as well.
 Thus the particle entered from a lead forgets where it came from.
 Such a process can be interpreted as a
 random-walk-like one and gives the same distribution function.
 It is interesting that the asymmetric random walk 
 with the probability $p\ne 0.5$ can be described by our model 
 with $X\ne 2$.
 Since the analytic form is not known in the asymmetric random walk,
 our result may be useful for understanding the result.
 It is also known that 
 the same distribution function appears in the problem of 
 the continuous-time quantum walk \cite{konno}.

\begin{figure}[tb]
\begin{center}
\includegraphics[width=0.9\columnwidth]{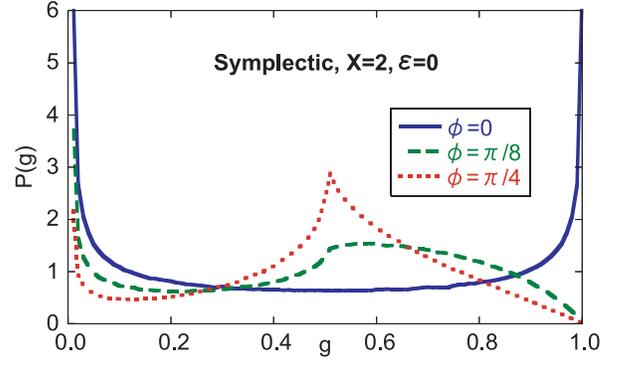}
\caption{Conductance distribution function of the symplectic system 
for several values of $\phi$.
We set $X=2$ and $\e=0$.
}
\label{spgp}
\end{center}
\end{figure}

 When the phase $\phi$ is finite, 
 $K_1$ does not belong to the member of the orthogonal ensemble
 and the results can depend on the choice of the ensembles.
 We numerically found that the orthogonal and unitary cases 
 are independent of $\phi$ and the result (\ref{pg0}) is kept unchanged.
 For the symplectic case, we found that the result depends on $\phi$
 and Eq.(\ref{pg0}) does not maintain anymore.
 The numerical result for $X=2$ and $\e=0$
 is shown in Fig.\ref{spgp}. 
 Remarkably, 
 all plotted curves give the averaged conductance $g=0.5$
 and the phase dependence appears only for the conductance fluctuations.
 We also see that plotted curves 
 has a nonanalytic point at around $g=0.5$, which implies 
 a nontrivial mechanism due to the phase coherent effect.
 It is not clear how this happens 
 and further study is needed to clarify the underlying mechanism.

\section{Dephasing}
\label{dephasing}

 In the Hamiltonian approach, 
 the dephasing effect can be modeled by 
 introducing the imaginary part to the energy
\be
 \e\to\e+\frac{i}{2\tau}.
\ee
 This method is equivalent with that of Ref.\cite{PEI}
 where the imaginary part of the Hamiltonian was introduced.
 In the supersymmetry method, this effect can be described 
 by the additional term of the sigma model \cite{PEI}
\be
 F_\tau = \frac{1}{\Delta\tau}\str \sigma\Lambda.
 \label{Ftau}
\ee
 This term makes the massless ``diffusion'' modes on 
 the saddle-point manifold massive and reduces the quantum fluctuations.
 See the Appendix \ref{calcond} for details.

 It is well known in the S matrix approach that 
 the dephasing effect can be described by 
 the B\"uttiker's voltage probe model \cite{Buttiker}.
 A fictitious voltage probe eliminating the phase coherence 
 is attached to the dot and is described by an enlarged S matrix.

 Brouwer and Beenakker showed 
 that the voltage-probe model at a certain limit 
 becomes equivalent to the imaginary-potential model and 
 found the modified Poisson kernel 
 in the random S matrix approach \cite{BB97}. 
 Here we investigate this limit using the imaginary-potential model.
 Since the dephasing effect to the regular dot is trivial, 
 we include the effect in the random dot only.

\begin{figure}[tb]
\begin{center}
\includegraphics[width=0.9\columnwidth]{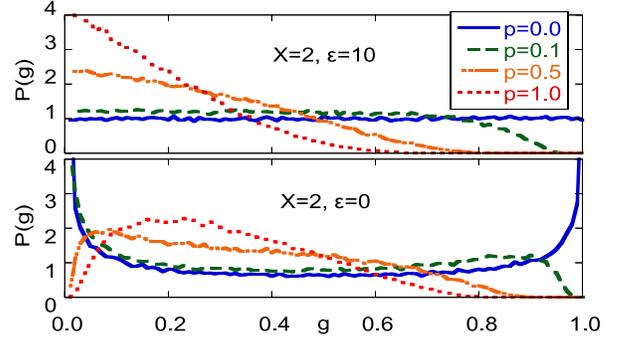}
\caption{Conductance distribution function for unitary system
 for several values of $p=1/\Delta\tau$.
}
\label{dpg}
\end{center}
\end{figure}

 In Fig.\ref{dpg}, the numerical results of the conductance 
 distribution function using the random Hamiltonian model are shown.
 We add the dephasing term, $p=1/\Delta\tau$ with 
 the phenomenological dephasing rate $\tau$, to the Hamiltonian.
 The matrix elements of the dot-lead coupling $w^{(2)}$ are 
 chosen randomly so that 
 there is no direct nonresonant reaction $a=0$. 
 The size of the random Hamiltonian is taken $10^2$ 
 and the averaging over $10^5$ samples is carried out. 

 As $p$ increases, the curve transforms into a single peak structure.
 The peak point corresponds to the mean part of the conductance $g_0$,
 which is close to zero for the upper graph and 0.25 for the lower one.
 We can conclude that the dephasing effect only affects
 the fluctuation part.
 We also confirmed that our numerical result based on 
 the random Hamiltonian 
 agrees with that of the random S matrix model in Ref.\cite{BB97}.

\section{Conclusions}
\label{conc}

 We have discussed an AB ring system with 
 regular and random cavities.
 We found that the quantum fluctuation effect plays 
 an important and crucial role
 and significantly affects the conductance.
 The main results are summarized as follows:
 (i) The averaged conductance is divided into two parts.
 The mean part has the Breit-Wigner resonant form renormalized
 by random effects.
 The quantum fluctuation part has an antiresonance form  
 where the quantum effects become minimal at the resonant point.
 (ii) For the orthogonal and symplectic ensembles, 
 the AB oscillations are found and the period of the oscillations
 is half a flux quantum.
 The positive (negative) magnetoconductance are obtained 
 for the orthogonal (symplectic) ensemble
 because of the multiple reflections inside of the ring.
 (iii) Depending on the parameter choice, 
 the Fano effect can be observed.
 This effect appears in the mean and fluctuation parts, respectively, 
 and a symmetric form is obtained for the total conductance. 
 (iv) The conductance distribution functions clearly show 
 the influence of strong fluctuations. 
 The distribution function at the resonant point, Eq.(\ref{pg0}),
 does not depend on the choice of the ensemble,
 which can be understood by the mode-locking mechanism.
 The form of the distribution function implies a relation to 
 the random walk problem.
 (v) The dephasing effect simulated by the imaginary-potential model
 reduces the fluctuation part only.

 The result of the averaged conductance in Fig.\ref{ge}
 shows that the total conductance as a function of the energy 
 takes a broad distribution.
 The form of the total conductance is determined by 
 the competition between the mean (\ref{gcl}) and 
 fluctuation (\ref{g1}) parts. 
 At large $X$ we can observe the antiresonance. 

 Separating the mean and fluctuation parts 
 is crucial to understand the obtained result.
 For example the Fano effect is found in both parts, 
 while the total conductance, the sum of them, becomes symmetric.
 We also found that the dephasing effect suppresses the fluctuation part, 
 which means that the cancellation is incomplete and 
 the asymmetric form can be obtained in a system with dephasing.

 The most striking result can be seen in 
 the calculation of the conductance distribution function.
 At the resonant point, the effective modes of the K matrix
 in Eq.(\ref{Kpara}) are locked 
 to those for the orthogonal ensemble.
 Only the orthogonal modes are amplified by 
 the multiple scattering through the ring and 
 we can find the ensemble-insensitive result.
 This result suggests a possibility of controlling
 the ensemble dependence of random systems 
 by the resonant singularity embedded in the systems.

 Our results for a coupled system show that 
 the nontrivial phenomena which are absent 
 in the single system can be observed in the hybrid system,
 which opens a new direction for theoretical and experimental studies
 of chaotic scattering.
 In this paper we only considered the regular-random coupled system.
 It is interesting to see more complicated systems such as 
 a random-random system and triple coupled cavities, and so on.
 A study of the series coupled random dot can be seen, e.g., 
 in Ref.\cite{IWZ}.
 To the best of our knowledge, there is no systematic study 
 on the parallel coupled system.
 It will be discussed in detail in a future publication \cite{AT}.
  
\section*{ACKNOWLEDGMENTS}

 We are grateful to D. Cohen and S. Iida for useful discussions.

\appendix*
\section{Calculation of the conductance}
\label{calcond}

 We calculate the conductance using 
 the nonlinear sigma model with unitary symmetry, Eq.(\ref{nls}).
 The first step to do is to represent the conductance 
 as an integral of the $\sigma$ matrix.
 This is the standard prescription discussed in detail in Ref.\cite{VWZ}
 and we have 
\be
 \langle |S_{12}|^2\rangle
 &=& \biggl<\str \left(k\frac{1+\Lambda}{2}\tilde{K}
 \frac{1}{1+i\Lambda\tilde{K}}\right)_{12} \no\\ 
 &&\times
 \str \left(k\frac{1-\Lambda}{2}\tilde{K}
 \frac{1}{1+i\Lambda \tilde{K}}\right)_{21}\biggr>_F \no\\
 & & +\biggl<\str \left(k\frac{1+\Lambda}{2}\tilde{K}
 \frac{1}{1+i\Lambda\tilde{K}}\right)_{11} \no\\
 &&\times
 \left(k\frac{1-\Lambda}{2}\tilde{K}
 \frac{1}{1+i\Lambda\tilde{K}}\right)_{22}\biggr>_F, \label{s12}
\ee
 where
 $k={\rm diag}(1,-1)$ in superspace, 
 $\langle\ \rangle_F$ denotes the integration over $\sigma$
 with the weight $e^{-F}$, and 
\be
 \tilde{K} &=& \frac{1}{E-E_1}\gamma_1-\frac{i\pi}{N\Delta}\gamma_2\sigma \no\\
 &=& \frac{1}{\e}\Phi-\frac{iX}{2}\sigma.
\ee
 In the second line we used $a=0$.

 The second step is to parametrize the supermatrix $\sigma$.
 We use \cite{Efetov} 
\be
 & & \sigma = U\sigma_0\bar{U}, \no\\
 & & \sigma_0=\bmat{cc} \cos\hat{\theta} & i\sin\hat{\theta} \\
 -i\sin\hat{\theta} & -\cos\hat{\theta} \emat_{\rm RA}, \no\\
 & & U = \bmat{cc} u & 0 \\ 0 & v \emat_{\rm RA},
\ee
 where 
\be
 & & \hat{\theta} = \bmat{cc} i\theta_B & 0 \\ 
 0 & \theta_F \emat_{\rm BF},
\ee
 and the integration range is given by 
 $0<\theta_B<\infty$ and $0<\theta_F<\pi$.
 $U$ includes the anticommuting Grassmann variables 
 and can be written as 
\be
 & & u=u_1u_2, \no\\
 & & u_1 = \exp\bmat{cc} 0 & i\eta \\ -i\eta^* & 0 \emat_{\rm BF}, \no\\
 & & u_2 = \bmat{cc} \mbox{e}^{i\varphi_1} & 0 \\ 
 0 & \mbox{e}^{i\varphi_2} \emat_{\rm BF},  \no\\
 & & v = \exp\bmat{cc} 0 & \kappa \\ -\kappa^* & 0 \emat_{\rm BF},
\ee
 where $\eta$ and $\chi$ are Grassmann variables and 
 the range of the real variables $\varphi_{1,2}$ 
 is given by $0<\varphi_{1,2}<2\pi$.
 The invariant measure of this parametrization is 
\be
 {\cal D}\sigma &=& Cd\theta_B d\theta_F 
 d\varphi_1 d\varphi_2
 d\eta d\eta^* d\kappa d\kappa^* \no\\
 & & \times
 \frac{\sinh\theta_B\sin\theta_F}{(\cosh\theta_B-\cos\theta_F)^2},
\ee
 where $C$ is the normalization constant.
 In this parametrization, we can write  
\be
 e^{-F} = \frac{\left[1-\frac{T_1}{2}(1-\cos\theta_F)\right]
 \left[1-\frac{T_2}{2}(1-\cos\theta_F)\right]}
 {\left[1+\frac{T_1}{2}(\cosh\theta_B-1)\right]
 \left[1+\frac{T_2}{2}(\cosh\theta_B-1)\right]}. \no\\
\ee

 The last step is to carry out the integrations.
 This calculation is cumbersome although it is a straightforward task.
 The first term in Eq.(\ref{s12}) includes the mean part $g_0$.
 It is easily obtained by substituting $\sigma=\Lambda$.
 The fluctuation correction is obtained from the integral 
\be
 & & \frac{1}{16}\int_1^\infty ds_1\int_{-1}^1ds
 e^{-F} \no\\
 & & \times
 \biggl|
 T_1\frac{1-\frac{T_1}{2}\left(1+\frac{X}{2}\right)}
 {\left[1+\frac{T_1}{2}(s_1-1)\right]
 \left[1-\frac{T_1}{2}(1-s)\right]} \no\\
 & & -T_2\frac{1-\frac{T_2}{2}\left[1+\frac{X}{2}
 +\frac{2}{X}\left(\frac{1}{\e^2}+\frac{i}{\e}\right)\right]}
 {\left[1+\frac{T_2}{2}(s_1-1)\right]
 \left[1-\frac{T_2}{2}(1-s)\right]}
 \biggr|^2. 
 \label{int1}
\ee
 In the same way, the second term in Eq.(\ref{s12}) 
 is reduced to 
\be
 & & \frac{1}{16}\int_1^\infty ds_1\int_{-1}^1ds
 \frac{1}{(s_1-s)^2}e^{-F} \no\\
 & & \times
 \Biggl\{
 \left[\frac{T_1}{1+\frac{T_1}{2}(s_1-1)}
 +\frac{T_2}{1+\frac{T_2}{2}(s_1-1)}\right]^2(s_1^2-1) \no\\
 & & 
 +\left[\frac{T_1}{1-\frac{T_1}{2}(1-s)}
 +\frac{T_2}{1+\frac{T_2}{2}(1-s)}\right]^2(1-s^2)
 \Biggr\}. \no\\
 \label{int2}
\ee
 We note that these expressions are obtained after integrating the
 Grassmann variables and changing the variables as 
 $s_1=\cosh\theta_B$ and $s=\cos\theta_F$.
 A careful manipulation is required 
 to carry out the remaining integrals.
 After lengthy calculations we can obtain Eq.(\ref{g1}).

 It is a straightforward task to include the dephasing effect
 described by the dephasing term (\ref{Ftau}).
 In the present parametrization, it can be written as
\be
 F_{\tau} = \frac{2}{\Delta\tau}(s_1-s),
\ee
 and is incorporated in the integrals 
 in Eqs.(\ref{int1}) and (\ref{int2}) 
 as $e^{-F_{\tau}}$.
 Although we do not show the analytical result explicitly, 
 it is not difficult to carry out the integrals.
 At the limit $|E-E_1|\to\infty$, 
 we can find the result of Ref.\cite{PEI}.



\begin{thebibliography}{99}
\bibitem{CS}
 For a recent review, see 
 Y.V. Fyodorov, T. Kottos, and H.-J. St\"ockmann,
 J. Phys. A {\bf 38}, 10433 (2005).

\bibitem{WD}
 E.P. Wigner, Ann. Math. {\bf 53}, 36 (1951);
 F.J. Dyson, J. Math. Phys. {\bf 3}, 140 (1962); 
 {\bf 3}, 157 (1962); {\bf 3}, 166 (1962).

\bibitem{Mehta}
 M.L. Mehta, {\it Random Matrices}, 3rd ed. (Academic, New York, 2004).

\bibitem{GMW}
 T. Guhr, A. M\"uller-Groeling, and H.A. Weidenm\"uller,
 Phys. Rep. {\bf 299}, 189 (1998).

\bibitem{MRWHG} 
 C.M. Marcus, A.J. Rimberg, R.M. Westervelt, P.F. Hopkins, and A.C. Gossard,
 Phys. Rev. Lett. {\bf 69}, 506 (1992).

\bibitem{Beenakker}
 C.W.J. Beenakker, Rev. Mod. Phys. {\bf 69}, 731 (1997).

\bibitem{Alhassid}
 Y. Alhassid, Rev. Mod. Phys. {\bf 72}, 895 (2000).

\bibitem{Hacken}
 G. Hackenbroich, Phys. Rep. {\bf 343}, 463 (2001).

\bibitem{ABG}
 I.L. Aleiner, P.W. Brouwer, and L.I. Grazman, 
 Phys. Rep. {\bf 358}, 309 (2002).

\bibitem{YHMS}
 A. Yacoby, M. Heiblum, D. Mahalu, and H. Shtrikman, 
 Phys. Rev. Lett. {\bf 74}, 4047 (1995).

\bibitem{KASKI}
 K. Kobayashi, H. Aikawa, A. Sano, S. Katsumoto, and Y. Iye, 
 Phys. Rev. B {\bf 70}, 035319 (2004).

\bibitem{GIA}
 Y. Gefen, Y. Imry, and M.Ya. Azbel, 
 Phys. Rev. Lett. {\bf 52}, 129 (1984).

\bibitem{KK}
 B. Kubala and J. K\"onig,
 Phys. Rev. B {\bf 65}, 245301 (2002); 
 {\bf 67}, 205303 (2003).

\bibitem{KAKI}
 K. Kobayashi, H. Aikawa, S. Katsumoto, and Y. Iye,
 Phys. Rev. Lett. {\bf 88}, 256806 (2002);
 Phys. Rev. B {\bf 68}, 235304 (2003).

\bibitem{NTA}
 T. Nakanishi, K. Terakura, and T. Ando, 
 Phys. Rev. B {\bf 69}, 115307 (2004).

\bibitem{VWZ}
 J.J.M. Verbaarschot, H.A. Weidenm\"uller, and M.R. Zirnbauer,
 Phys. Rep. {\bf 129}, 367 (1985).

\bibitem{pker}
 L.K. Hua, {\it Harmonic Analysis of Functions of Several Complex 
 Variables in the Classical Domains} 
 (American Mathematical Society, Providence, 1963);
 P.A. Mello, P. Pereyra, and T.H. Seligman,
 Ann. Phys. (N.Y.) {\bf 161}, 254 (1985);
 P.A. Mello and N. Kumar, 
 {\it Quantum Transport in Mesoscopic Systems}
 (Oxford University Press, Oxford, 2004).

\bibitem{PEI}
 V.N. Prigodin, K.B. Efetov, and S. Iida, 
 Phys. Rev. Lett. {\bf 71}, 1230 (1993);
 Phys. Rev. B {\bf 51}, 17223 (1995).

\bibitem{ISS}
 F.M. Izrailev, D. Saher, and V.V. Sokolov, 
 Phys. Rev. E {\bf 49}, 130 (1994).

\bibitem{randomS}
 H.U. Baranger and P.A. Mello,
 Phys. Rev. Lett. {\bf 73}, 142 (1994);
 R.A. Jalabert, J.-L. Pichard, and C.W.J. Beenakker,
 Europhys. Lett. {\bf 27}, 255 (1994).

\bibitem{BB94}
 P.W. Brouwer and C.W.J. Beenakker,
 Phys. Rev. B {\bf 50}, R11263 (1994).

\bibitem{Brouwer}
 P.W. Brouwer, Phys. Rev. B {\bf 51}, 16878 (1995).

\bibitem{FS} 
 Y.V. Fyodorov and H.-J. Sommers, J. Math. Phys. 
 {\bf 38}, 1918 (1997); 
 J. Phys. A {\bf 36}, 3303 (2003).

\bibitem{Stockmann}
 H.-J. St\"ockmann, {\it Quantum Chaos: An Introduction}
 (Cambridge University Press, Cambridge, U.K., 1999).

\bibitem{HZOAA}
 S. Hemmady, X. Zheng, E. Ott, T.M. Antonsen, and S.M. Anlage, 
 Phys. Rev. Lett. {\bf 94}, 014102 (2005); 
 S. Hemmady, X. Zheng, J. Hart, T.M. Antonsen, Jr., E. Ott, 
 and S.M. Anlage, Phys. Rev. E {\bf 74}, 036213 (2006).

\bibitem{RLBKS}
 S. Rotter, F. Libisch, J. Burgd\"orfer, U. Kuhl, and H.-J. St\"ockmann, 
 Phys. Rev. E {\bf 69}, 046208 (2004).

\bibitem{KLAA}
 V.E. Kravtsov, I.V. Lerner, B.L. Altshuler, and A.G. Aronov,
 Phys. Rev. Lett. {\bf 72}, 888 (1994).

\bibitem{micolich}
 A.P. Micolich {\it et al.}, 
 Phys. Rev. Lett. {\bf 87}, 036802 (2001).

\bibitem{SI}
 P.G. Silvestrov and Y. Imry, Phys. Rev. Lett. {\bf 85}, 2565 (2000).

\bibitem{BHZ}
 E. Br\'ezin, S. Hikami, and A. Zee,
 Phys. Rev. E {\bf 51}, 5442 (1995).

\bibitem{CWB}
 A.A. Clerk, X. Waintal, and P.W. Brouwer, 
 Phys. Rev. Lett. {\bf 86}, 4636 (2001).

\bibitem{Efetov} K.B. Efetov, Adv. Phys. {\bf 32}, 53 (1983); 
 {\it Supersymmetry in Disorder and Chaos} 
 (Cambridge University Press, Cambridge, U.K., 1997).

\bibitem{HPMBDH}
 A.G. Huibers, S.R. Patel, C.M. Marcus, P.W. Brouwer, 
 C.I. Duru\"oz, and J.S. Harris, Jr., 
 Phys. Rev. Lett. {\bf 81}, 1917 (1998).

\bibitem{TA}
 K. Takahashi and T. Aono, Phys. Rev. B {\bf 74}, 041311(R) (2006).

\bibitem{BB97}
 P.W. Brouwer and C.W.J. Beenakker, Phys. Rev. B {\bf 55}, 4695 (1997).

\bibitem{Efetov2}
 K.B. Efetov, Phys. Rev. Lett. {\bf 74}, 2299 (1995).

\bibitem{AAS}
 B.L. Al'tshuler, A.G. Aronov, and B.Z. Spivak, 
 Pis'ma Zh. Eksp. Teor. Fiz. {\bf 33}, 101 (1981) 
 [JETP Lett. {\bf 33}, 94 (1981)]; 
 D.Yu. Sharvin and Yu.V. Sharvin, 
 {\it ibid.} {\bf 34}, 285 (1981)
 [{\it ibid.} {\bf 34}, 272 (1981)]; 
 B.L. Al'tshuler, A.G. Aronov, B.Z. Spivak, D.Yu. Sharvin, 
 and Yu.V. Sharvin, {\it ibid.} {\bf 35}, 476 (1982) 
 [{\it ibid.} {\bf 35}, 588 (1982)].

\bibitem{AAS2}
 J.P. Carini, K.A. Muttalib, and S.R. Nagel, 
 Phys. Rev. Lett. {\bf 53}, 102 (1984); 
 M. Murat, Y. Gefen, and Y. Imry, Phys. Rev. B {\bf 34}, 659 (1986);
 see also Y. Imry, {\it Introduction to Mesoscopic Physics} 
 (Oxford University Press, Oxford, 1997).

\bibitem{weakl}
 For a review, see 
 S. Chakravarty and A. Schmid, Phys. Rep. {\bf 140}, 193 (1986).

\bibitem{Fano}
 U. Fano, Phys. Rev. {\bf 124}, 1866 (1961).

\bibitem{Heermann}
 See for example, 
 D.W. Heermann,
 {\it Computer Simulation Methods in Theoretical Physics} 
 (Springer-Verlag, Berlin, 1986)

\bibitem{feller}
 W. Feller,
 {\it An Introduction to Probability Theory and Its Applications}, 
 3rd ed. (Wiley, New York, 1968).

\bibitem{konno}
 N. Konno, Phys. Rev. E {\bf 72}, 026113 (2005).

\bibitem{Buttiker}
 M. B\"uttiker, Phys. Rev. B {\bf 33}, 3020 (1986);
 IBM J. Res. Dev. {\bf 32}, 63 (1988).

\bibitem{IWZ}
 S. Iida, H.A. Weidenm\"uller, and J.A. Zuk,
 Phys. Rev. Lett. {\bf 64}, 583 (1990);
 Ann. Phys. (N.Y.) {\bf 200}, 219 (1990).

\bibitem{AT}
 T. Aono and K. Takahashi (unpublished).

\end{thebibliography}
\end{document}